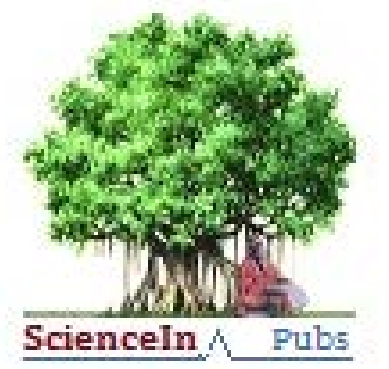

# Automated brain tumor detection in MRI images using CNN and ResNet architectures

Annapurna V K[1], Asha N[2], K Paramesha*[3], Shabana Sultana[1], Kirankumar Humse[4]

*[1]Department of Computer Science and Engineering, The National Institute of Engineering, Mysuru, Karnataka, India. [2]Department of MCA, The National Institute of Engineering, Mysuru, Karnataka, India. [3]Department of Computer Science and Engineering, Vidyavardhaka College of Engineering, Mysuru, Karnataka, India. [4]Department of ECE, Vidyavardhaka College of Engineering, Mysuru, Karnataka, India.*

## ABSTRACT

Deep learning has shown significant potential in medical image analysis, particularly for disease detection using MRI scans. Accurate and early diagnosis of brain tumors remains challenging due to the complexity of brain structures and reliance on manual interpretation. This work presents an automated deep learning–based approach for brain tumor detection from MRI images using Convolutional Neural Networks and Residual Networks. Transfer learning is applied with two pretrained architectures, ResNet18 and ResNet50, to classify MRI scans into tumor and non-tumor categories. Experiments are conducted on a dataset of 3,929 brain MRI images, evaluating the impact of model depth and fine-tuning strategies. The results show that ResNet18 achieves a higher accuracy of 97% compared to 96% for ResNet50, demonstrating better generalization on limited medical data. The proposed framework enables fast, accurate, and cost-effective brain tumor detection, supporting early diagnosis and clinical decision-making.

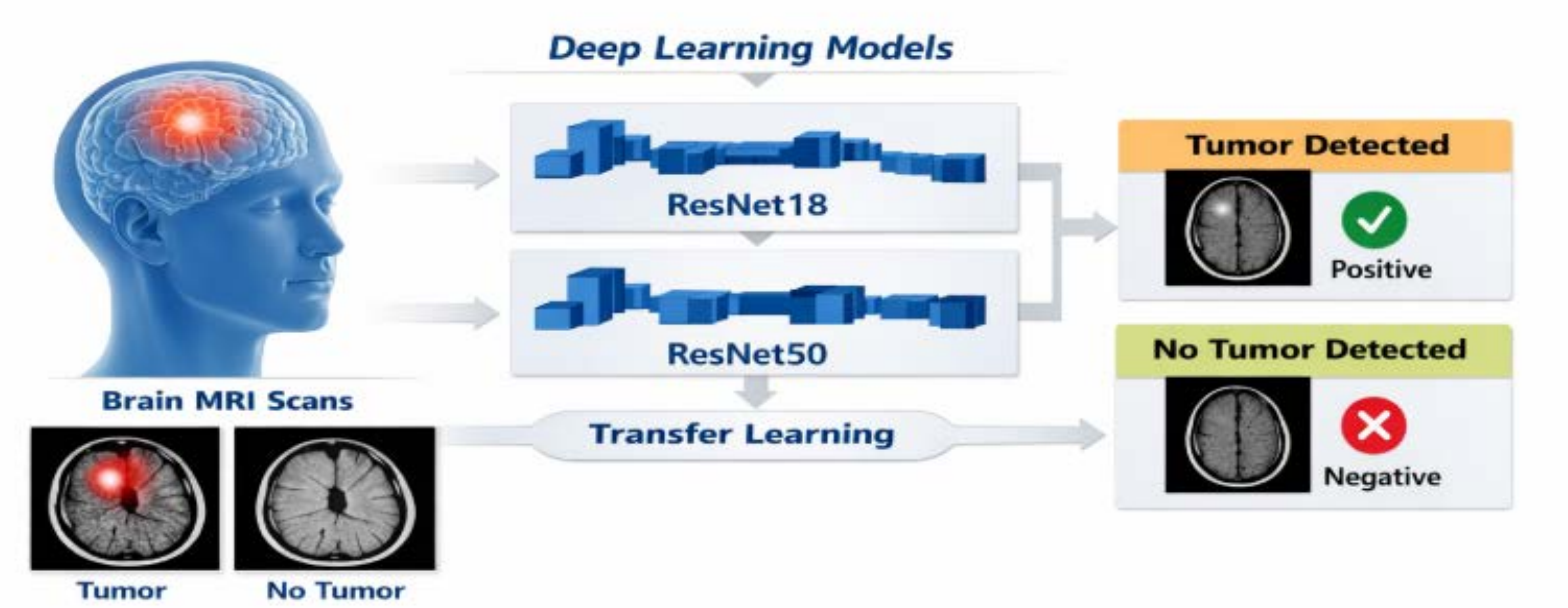

## INTRODUCTION

Brain tumors are such neurological diseases that can cause extreme morbidity and mortality when it becomes late. The early correct diagnosis is significant in the selection of suitable treatment modalities and improving patient survival rate in a very significant manner. In recent years, Magnetic Resonance Imaging (MRI) has emerged as the most desirable imaging modality in diagnosing brain tumors because it offers high-resolution images with high-soft tissue contrast making clinicians able to see the complex structures of the brain and the pathological abnormalities. MRI imaging has been widely utilized in the diagnosis of the presence, size, location, and progression of tumours, and is therefore an indispensable aspect of neuro-oncological practice.[1,2]

Although MRI has its benefits, radiologists still have a hard and demanding time of interpreting the brain scans manually. Diagnostic process could be rather subjective and inconsistent among experts, which results in inter-observer variability and possible diagnostic discrepancies. Also, recent advancements in medical imaging technologies have put an enormous strain on clinical practitioners, which exposes them to the danger of overlooking and late diagnosis. These hurdles highlight the need to have effective diagnostic tools that are reliable and automated and can help clinicians with more accurate, consistent, and quick detection of brain tumors to help improve clinical decisions, as well as overall healthcare efficiency.[3,4]

The MRI images are most often subject to expert interpretation to schedule the traditional diagnosis of brain tumors, which is not only a labour-consuming task but also prone to human error, particularly in low-contrast or small tumours. Traditional methods in machine learning have tended to rely on handcrafted features, restraining their capacity to generalize to different datasets and imaging settings. Furthermore, the growing need to diagnose fast in a clinical environment is leading to the need to have automated systems that will produce high-quality accuracy with minimum human intervention. Thus, there is an urgent necessity of an effective and automated deep learning-based model that could correctly identify the brain MRI scans as tumor-positive or tumor-negative.

*Corresponding Author: K Paramesha, Department of CSE, Computer Science and Engineering, Vidyavardhaka College of Engineering, Mysuru, Karnataka, India
Email: Paramesha.k@vvce.ac.in

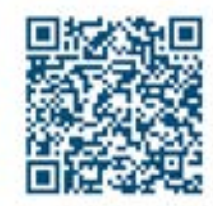

The latest developments in the field of deep learning, especially Convolutional neural networks (CNNs), have shown impressive performance in the areas of medical image analysis, including the classification of diseases, the localization of objects, and their segmentation. Deep learning models are able to automatically learn discriminative features on raw image data, which does not require handcrafted features. Among them, Residual Networks (ResNets) have demonstrated better performance due to the ability to resolve the vanishing gradient issue as well as train deeper networks.

Driven by the increased use of artificial intelligence in healthcare and its success as an automated medical imaging device, this paper examines the use of CNN and ResNet architectures to detect brain tumors using MRI scans automatically. The presence of ready-to-use models and transfer learning methods also encourage the application of deep residual networks to start with high quality with a small number of labeled medical samples. Also, a comparative study of ResNet18 and ResNet50 can give an insight into the trade-offs between model depth, performance and computational complexity. The key contributions of this work are summarized as follows:

- Development of an automated deep learning framework for brain tumor detection from MRI images using CNN and ResNet architectures.
- Implementation and evaluation of ResNet18 and ResNet50 models to classify MRI scans into tumor-positive and tumor-negative categories.
- Application of transfer learning techniques, including weight freezing and fine-tuning, to improve model performance and training efficiency.
- Comprehensive comparative performance analysis of the two residual network architectures using a dataset of 3929 brain MRI images.
- Demonstration of a cost-effective and scalable diagnostic approach that has the potential to assist radiologists and improve early detection of brain tumors.

## RELATED WORKS

Gokmen[5] has provided a comprehensive overview of artificial intelligence methods used to segment brain tumors using MRI in 2019-2024. The research systematically analyzed both the classic machine learning techniques and modern deep learning models such as CNNs, U-Net versions, and residual networks. The author highlighted the increasing trend in using deep learning because of its high feature extracting abilities but also noted that there are still persistent issues, including scarce labeled data, class-imbalanced samples, computational costs, and poorer interpretations of the model.[5] The explainability of deep learning-based brain tumor detection is a critical issue that Guluwadiresolved through the incorporation of ResNet50 and Grad-CAM visualization. This method allowed determining salient areas that affected model predictions, which enhanced transparency and clinical trust.[6] Hussain et al. also contributed to the discipline by suggesting the next-generation neural network architecture to provide brain tumor segmentation and classification using MRI. Their contribution showed better automation, robustness, and scalability, which show that deep learning systems have potential to help clinicians in neuro-oncology workflows.[7]

To enhance the current state of automatic segmentation, Saifullah et al. introduced an enhanced automatic segmentation framework that integrates a ResNet50 encoder into a U-Net with high accuracy.[8] A comparative study of deep learning methods against automated brain tumor detection, by Kumar and Sachar obtained affirmative results by showing that CNN-based models have invariably higher performance compared to conventional handcrafted feature-based systems.[9] Kia et al. proposed a novel ensemble fusion approach that combines VGG16, MobileNet, EfficientNet, AlexNet, and ResNet50 and showed that a fusion of multiple models outperforms single models in terms of the robustness and accuracy of classifications.[10] On the same note, Rao et al. compared and estimated various pre-trained CNN structures and demonstrated that transfer learning is effective in reducing training time and increasing classification performance.[11] Yang et al. introduced BrainCNN, a specialized CNN network that is specifically trained on automated brain tumor grading of MRI images. The model demonstrated high grading accuracy through task-specific hierarchical features learning.[12] Puttegowda et al. generalized machine learning to the analysis of breast cancer mammograms and showed how high-level learning models could be applied with various medical imaging modalities.[13] Rani et al. have introduced a U-Net ensemble architecture based on ResNet, which demonstrated greater resilience and precision in the detection of brain tumor in heterogeneous MRI databases.[14] Conversely, Parameshachari and Sunil Kumar examined an SVM-based brain tumor detector, which, though was effective, had drawbacks compared to the deep learning-based systems.[15]

Benbakreti et al.examined the effect of data augmentation on CNN and pre-trained models, revealing that augmentation does contribute a lot to the generalization and overfitting in brain tumor detection.[16] The comparative analysis of deep learning networks by Mahmud et al.was very thorough, and the authors stressed that more profound residual networks are more efficient in dealing with intricate tumor patterns in MRI images.[17] Ali and Agrawal used VGG19 and ResNet101 to perform automated brain tumor segmentation, with great accuracy of segmentation as a result of deep hierarchical feature learning.[18] Prabhavathi and Anandararaju concentrated on the issue of medical image security by an effective encryption algorithm in telemedicine application that discussed the issue of data privacy in relation to AI-based healthcare systems.[19] Saranya and Praveena proposed an optimized architecture based on YOLOv5 to detect and classify brain tumors in real-time, showing that this approach could be applied in time-critical cases of diagnosing the disease.[20] Karthik et al.used machine learning to detect epileptic seizures through improved EEG signal processing procedures, as the application of machine learning in the diagnosis of neurological disorders, not medical imaging.[21]

Aamir et al. have put forward an automated deep learning system in brain tumor classification using MRI with high diagnostic performance using optimized CNN architectures.[22] The paper by Sudheesh et al. advances the wider application of deep learning in the field of neurological care by creating a neuroimaging-based deep learning model to predict Alzheimer disease.[23] Manjunath et al. have shown that deep learning models that are fine-tuned are highly effective in enhancing the accuracy of brain tumor classification based on transfer learning.[24] Kaur and Mahajan presented an optimized ResNet152 with transfer learning to detect brain tumors and it demonstrated a high level of robustness and accuracy.[25] Natesh et al. combined machine learning methods and external risk factors to improve detection of diabetes, demonstrating the ability of AI to be used in healthcare analytics.[26] A CNN-based automated brain tumor diagnosis framework was created by Chaudhary et al., which once again confirms the usefulness of deep learning in MRI-based diagnosis.[27] Raksha et al. explored the use of deep learning in early detection and staging of Alzheimer diseases and found out that diagnostic reliability improved.[28] Kumar et al. used deep learning algorithms to diagnose learning disorders in children with autism spectrum disorder at an early stage, which illustrates that AI can be used in cross-domain medical diagnostics.[29] Vijay et al. tested the chaperoned machine learning mechanisms to detect epileptic seizures based on multichannel EEG signals showing effective reliability in both patient-process-specific and generalized context.[30]

Based on the literature review, it is clear that CNN- and ResNet-based networks are the best at brain tumor detection and segmentation because of their high feature learning abilities. Although explainable AI, ensemble learning and real-time detection frameworks have achieved substantial performance improvement, the challenges of optimality of model depth, computational efficiency, and uniform interpretability have not yet been solved. These loopholes have inspired the current study, which gives a comparative analysis of ResNet18 and ResNet50 networks with the help of transfer learning to strike a balance between accuracy, efficiency, and clinical usability.

## PRELIMINARIES

The section introduces the basic ideas and methods, which the proposed brain tumor detection framework is based on. It gives a brief introduction to several important elements of deep learning applied in this paper image segmentation, Convolutional Neural Networks (CNNs), Residual Networks (ResNets), and transfer learning. Knowledge of these concepts is the key to comprehend the methodology and experimental analysis introduced in the following sections.

### IMAGE SEGMENTATION

Image segmentation refers to the process of labeling every pixel of an image with a class label, which allows one to understand the image on a pixel-by-pixel basis and extract information. Its main applications include medical image recognition and localization, as well as autonomous systems, among others. The neural networks in deep learning-based segmentation are trained to produce pixel-wise masks indicating pixel regions of interest.

### CONVOLUTIONAL NEURAL NETWORK

Convolutional Neural Network (CNN) is a type of neural network used to analyze visual data, which is one of the most efficient to use in image classification, object detection, and localization. CNNs can be trained to learn hierarchical features representations automatically on raw images which traditional networks like multilayer perceptrons are unable to do effectively. An average CNN is made of convolutional and pooling feature extraction layers, and fully connected final classification layers. Figure 1 illustrates the typical architecture of a Convolutional Neural Network (CNN.

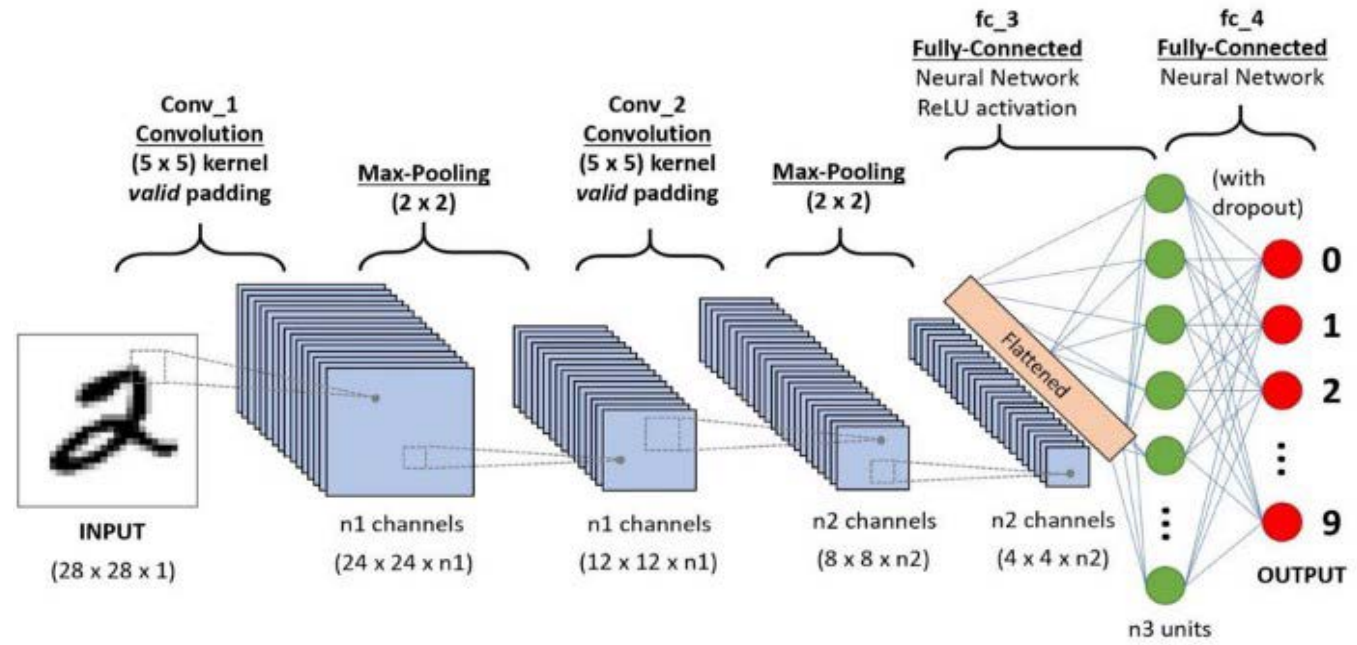

**Figure1.** Convolutional Neural Network Architecture

### RESIDUAL NETWORK

The residual Networks (resnets) are a type of deep neural networks that add skip connections to help solve the problem of the vanishing gradient with very deep networks. These shortcut connectors permit gradients to traverse identity mappings, and thus can be trained on deeper networks without information loss. The ResNet architectures consist of residual blocks that carry convolutional layers, ReLU activation, and batch normalization, after which there are global average pooling and Softmax classifier. The design enhances learning in features, stability and performance of the network. Figure 2 illustrates the structure of a Residual Network (ResNet), where residual (skip) connections allow the input of a layer to be added directly to its output.

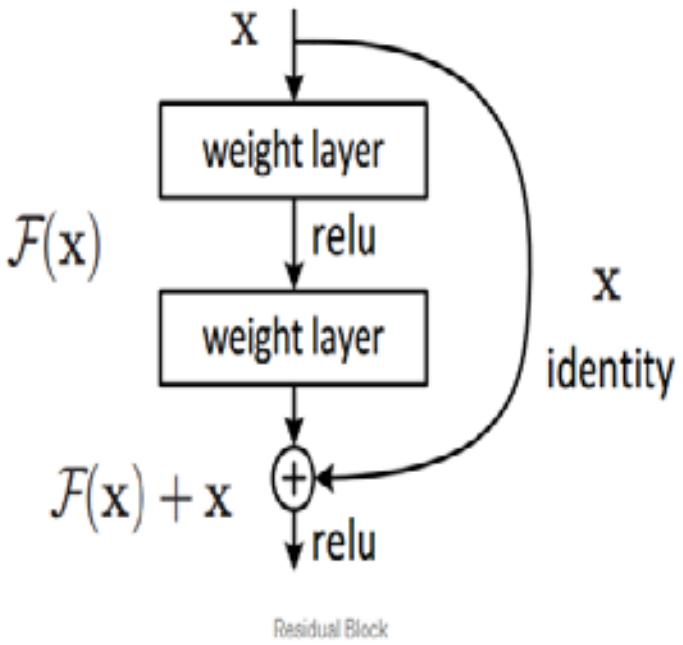

**Figure 2.** Residual Network

### TRANSFER LEARNING

Transfer learning is a methodology, where the information that a trained model has obtained on a large set of data is reapplied to a similar target task. This method proves effective when there are few labeled data or there is variation in the distribution of data across domains. Transfer learning saves a lot of time in training networks and enhances the performance of model, by fine-tuning or freezing the pretrained network layers. It breaks the constraints of learning individually since it allows the model to use prior knowledge and hence is very useful with highly complicated problems including medical image analysis.The fact that most models solve extremely difficult problems and require a great deal of data is one of the most significant applications of deep learning. As it is able to do so, therefore it removes issues of labeling data typically done on supervised learning problems in machine learning. Figure 3 illustrates the transfer learning framework, where a pretrained deep neural network is used as a feature extractor or fine-tuned for a target task.

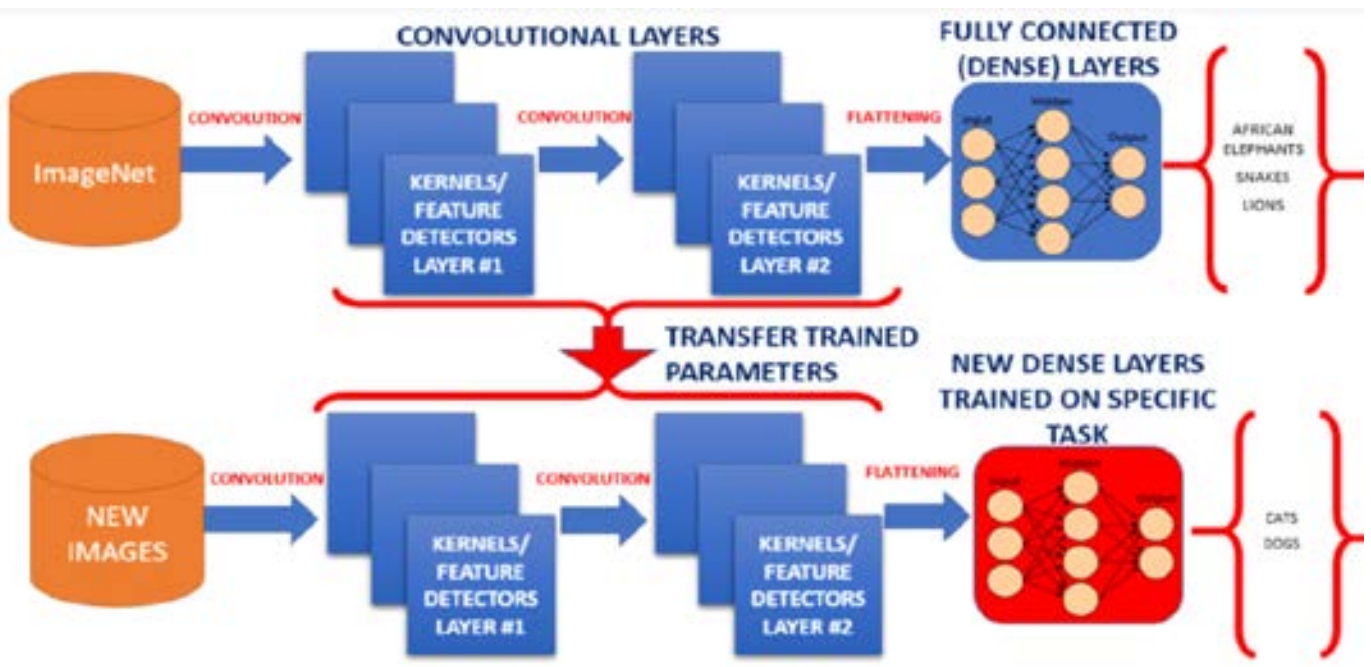

**Figure 3.** Transfer Learning Architecture

## PROPOSED METHODOLOGY

The proposed framework in the figure 4 presents an automated approach for brain tumor detection using MRI images based on deep learning techniques. Initially, the brain MRI dataset containing tumor and non-tumor images is subjected to image preprocessing, which includes resizing, normalization, and data augmentation to enhance image quality and improve model generalization. The processed dataset is then divided into training and testing sets. Transfer learning is employed using pre-trained ResNet18 and ResNet50 models to leverage learned feature representations. The networks perform deep feature extraction through residual CNN blocks, followed by fully connected layers with activation and dropout for effective classification. A Softmax classifier is used to categorize MRI scans into tumor or non-tumor classes. Finally, the system evaluates performance using accuracy metrics and compares the effectiveness of ResNet18 and ResNet50, identifying the optimal architecture for brain tumor detection.

The proposed deep learning models were trained and assessed using the Brain MRI scan images. Network architectures were used with residual Networks to categorize MRI scans as tumor and non-tumor. ResNet18 and ResNet50 are two pretrained models that were implemented and compared. The transfer learning techniques involved tuning the important hyperparameters by (a) freezing the initial convolutional layers and training only the newly added fully connected layers with initial random weights and (b) fine-tuning the entire network with pretrained weights through low learning rate to maintain the earlier learnt features and adapt to the new data.

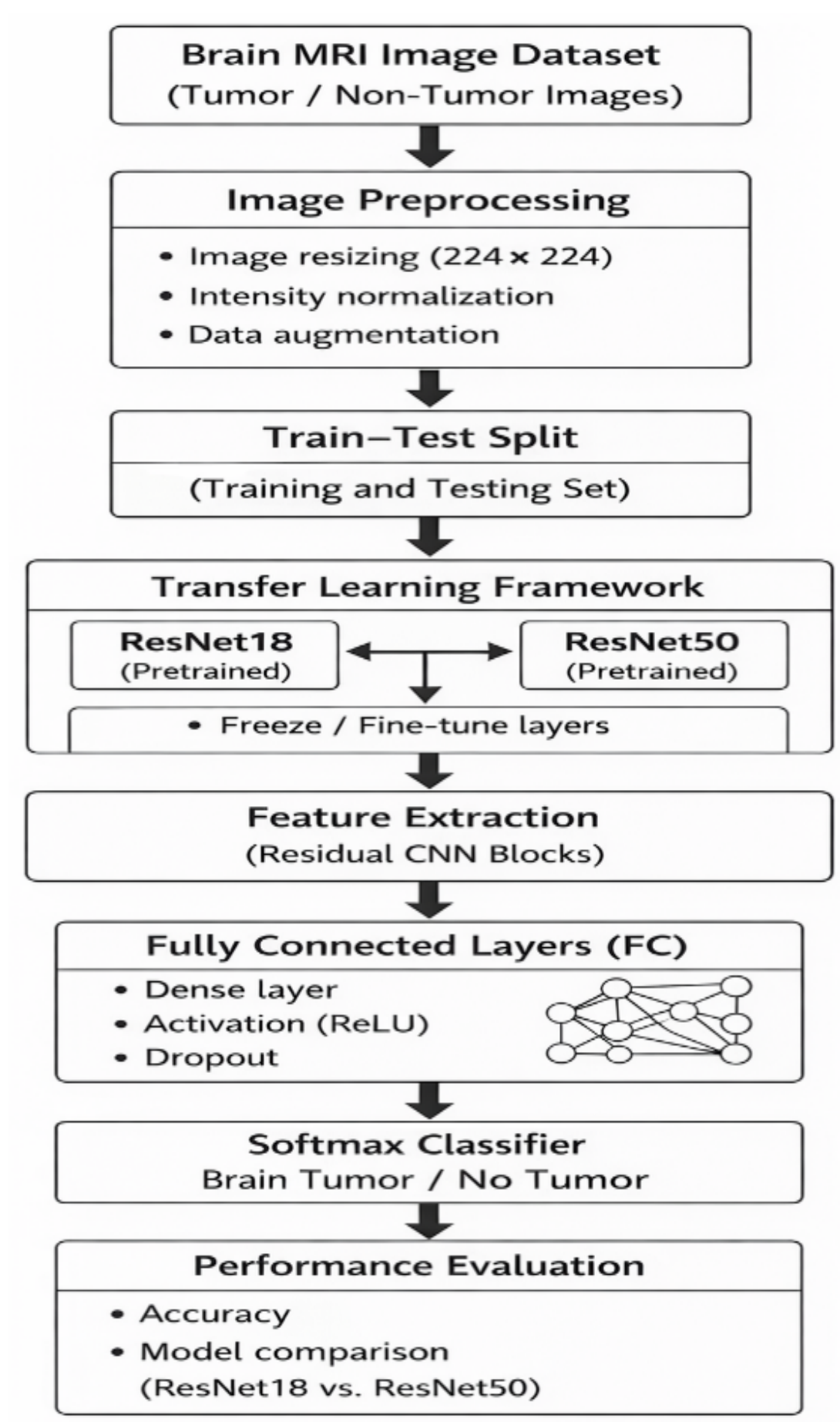

**Figure 4.** Block diagram of the Proposed System

### Dataset

The dataset will include 3,929 brain MRI scans which were classified by clinical experts as tumor or non-tumor cases. All the pictures were turned to PNG format and applied to the model training and assessment. Two deep learning models were created using the ResNet18 and ResNet50 architecture and important hyperparameters were varied in order to compare them. The data was divided into training and testing data, where 85 percent of the images were used as training data and the rest 15 percent of the images were used as testing data.

### Image Preprocessing

Before model training, MRI images are preprocessed so that they are consistent and their learning performance enhanced. All the images are downsampled to 224 × 224 pixels to fit the pretrained ResNet models. Normalization of pixel intensities is done in order to improve training convergence, data augmentation methods, such as flipping and rotation are employed to improve the data diversity, minimize overfitting, and enhance the model generalization.

### Train–Test Split

The preprocessed dataset is divided into two subsets: a training set and a testing set. The training set is used to learn the parameters of the model, while the testing set evaluates its performance on unseen data. This separation ensures unbiased evaluation and helps in assessing the real-world applicability of the proposed model.

**Transfer Learning Framework (ResNet18 and ResNet50)**

To improve classification performance and reduce training time, a transfer learning approach is adopted using pretrained ResNet18 and ResNet50 models. These models are pretrained on the ImageNet dataset and already contain rich feature representations. In this framework, either the early layers are frozen to retain learned features or selectively fine-tuned to adapt to brain MRI data. Both ResNet18 and ResNet50 are employed to analyze the impact of network depth on classification accuracy. Figure 5 illustrates the architecture of the ResNet50 (Residual Network with 50 layers) model, which is a deep convolutional neural network widely used for image classification tasks. ResNet50 is designed to overcome the limitations of very deep networks, particularly the vanishing gradient problem, by introducing residual (skip) connections.

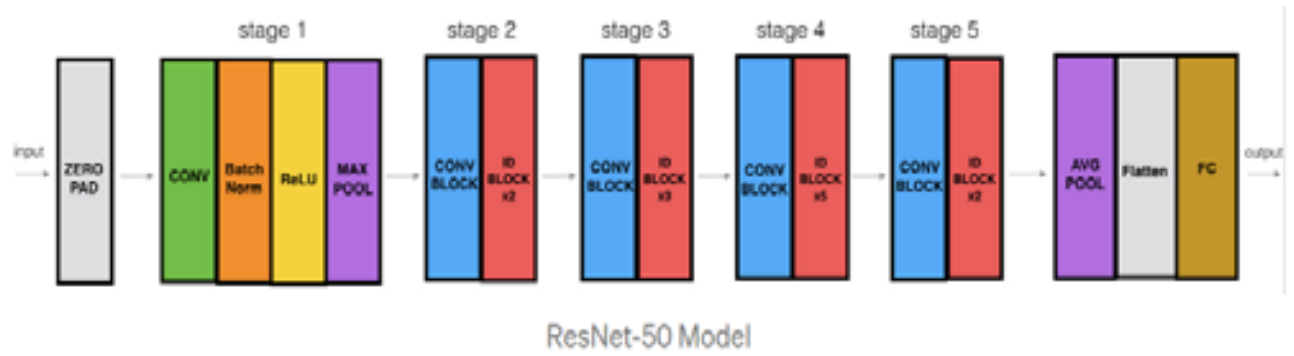

**Figure 5.** ResNet50 Model

Figure 6 illustrates the overall decision-making process of the proposed brain tumor detection system using a ResNet-based deep learning classifier. The input to the system is a brain MRI scan, which is first provided to the pretrained ResNet deep learning classifier model. The model automatically extracts high-level discriminative features from the MRI image using residual convolutional layers and analyzes structural and intensity variations associated with tumor presence. Based on the learned features, the classifier produces a binary decision. If tumor-related patterns are detected in the MRI scan, the system outputs "YES", indicating the presence of a brain tumor. Otherwise, it outputs "NO", indicating a non-tumor case. This framework enables automated, fast, and reliable brain tumor detection, reducing manual effort and supporting clinical decision-making.

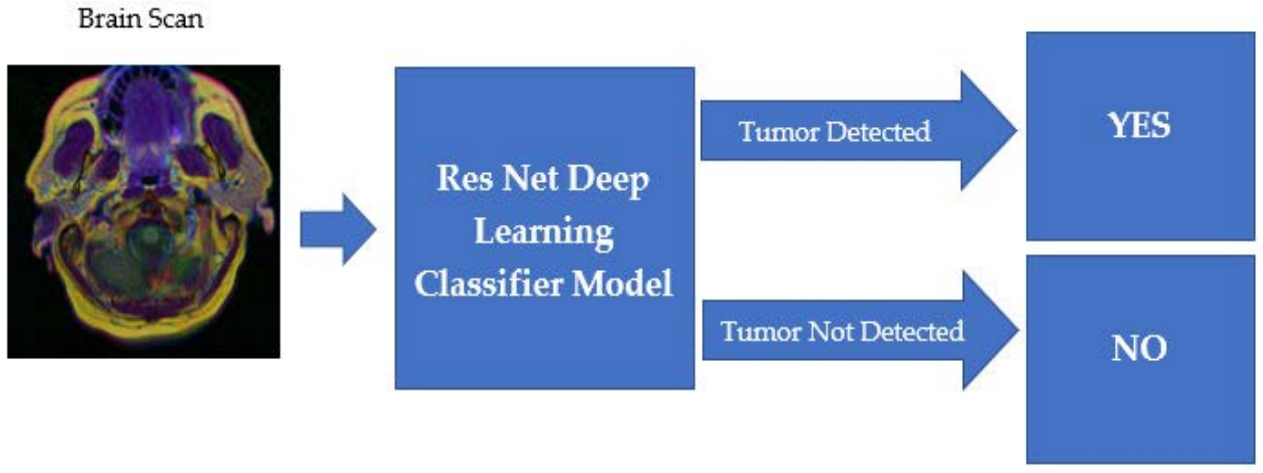

**Figure 6**. ResNet-Based Brain Tumor Detection and Classification Framework

**Feature Extraction (Residual CNN Blocks)**

The pretrained ResNet architectures act as powerful feature extractors. The residual convolutional blocks automatically learn hierarchical features such as edges, textures, shapes, and complex tumor patterns from MRI images. Skip connections in ResNet help mitigate the vanishing gradient problem, enabling efficient training of deep networks and preserving important spatial information.

**Fully Connected Layers (FC)**

The extracted deep features are passed to fully connected layers for final representation learning. These layers include dense neurons followed by ReLU activation to introduce non-linearity. Dropout regularization is applied to prevent overfitting by randomly disabling neurons during training. This stage transforms learned features into a format suitable for classification.

**Softmax Classifier (Brain Tumor / No Tumor)**

The final layer of the network is a Softmax classifier that assigns probabilities to each class. Based on the highest probability score, the MRI image is classified as either brain tumor or no tumor. Softmax ensures normalized probability distribution and supports clear decision-making for binary classification.

## RESULT AND DISCUSSION

To measure the effect of model architecture, preprocessing, and hyperparameter tuning on classification performance, a set of experiments were carried out. The preliminary findings revealed that ResNet18 outperformed ResNet50 by about 1 percent since its fewer parameters lessened its overfitting to the relatively small dataset. The accuracy of ResNet18 was 96 percent after the optimisation of hyperparameters, and the accuracy of resnet50 was 94 percent. The use of image normalization resulted in a performance drop of ResNet18 but slightly enhanced the results of ResNet50. More experiments showed that random horizontal flipping enhanced generalization, specifically in ResNet18. The highest overall performance was achieved by using random horizontal flipping followed by normalization where ResNet18 and ResNet50 obtained the highest accuracy of 97 and 96 respectively using the best training parameters.

**Accuracy**

The most common evaluation measure in classification problems is accuracy, which is a measure of the percentage of correctly classified examples of all of its predictions. It shows the degree of overall correctness of the model. The equation for accuracy as defined in (1).

$$Accuracy = \frac{TP + TN}{TP + TN + FP + FN} \quad (1)$$

**Precision**

Precision is used to determine the accuracy of positive predictions, i.e. the number of true positives of all positive cases that were predicted to be true. It is especially needed when it is crucial to reduce the number of false positives. The precision equation (2) is written as follows.

$$Precision = \frac{TP}{TP + FP} \quad (2)$$

**Recall**

Recall or sensitivity or true positive rate is used to determine the capability of the model to identify all the actual positive cases. It is particularly applicable in medical diagnosis, where false negative may be very grave. The equation (3) for recall can be expressed as follows.

$$Recall = \frac{TP}{TP + FN} \quad (3)$$

### F1 Score

F1-score is a composite of precision and recall into a single measure through the harmonic mean of the two measures. It offers an equalized assessment, especially in cases where there is class imbalance or where false positive as well as the false negative is equally significant.The equation (4) for calculating the F1 score is as follows.

$$F1\ Score = \frac{2 * \text{Precision} * \text{Recall}}{\text{Precision} + \text{Recall}} \quad (4)$$

The figure 7 highlights the comparative performance of the two transfer learning models. The superior performance of ResNet18 can be attributed to its lower number of parameters, which reduces overfitting and improves learning efficiency when trained on a relatively small dataset. The ResNet18 model achieves a higher accuracy of 97% compared to 96% obtained by ResNet50.

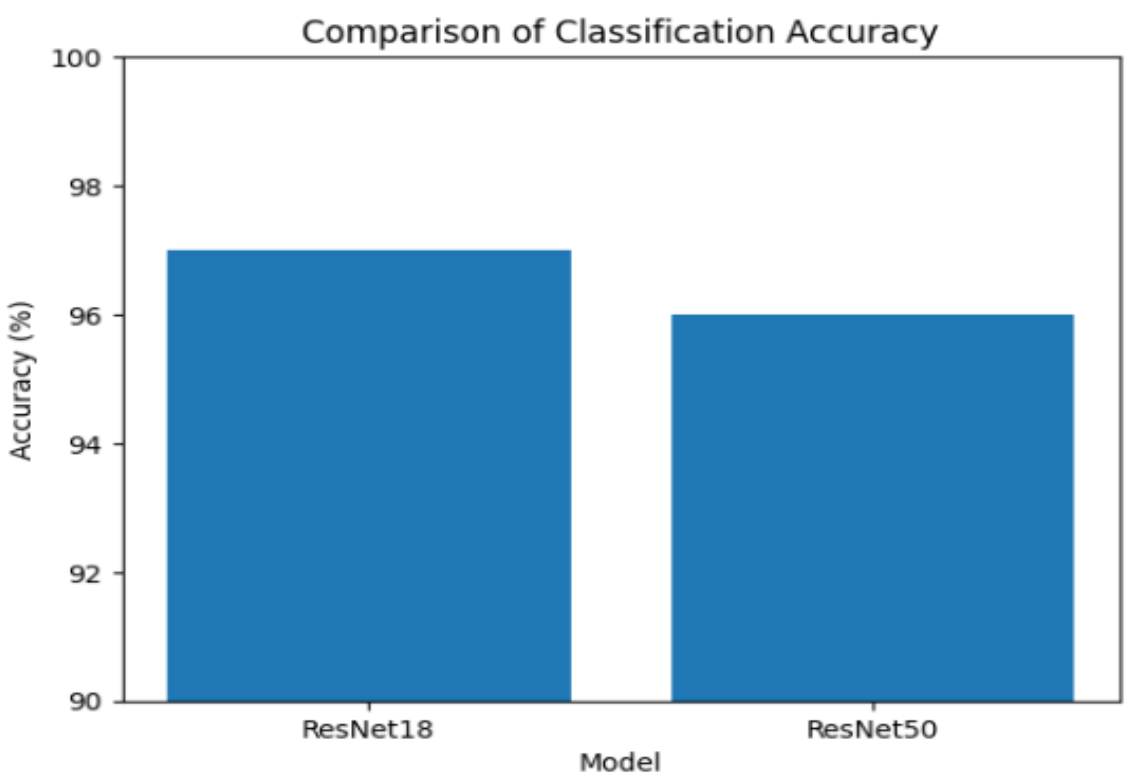

**Figure 7.** Comparison of classification accuracy between ResNet18 and ResNet50 models

Table 1 provides a comparative study of the effects of various preprocessing measures, i.e., the image resizing, intensity normalization, and random horizontal flipping as data augmentation, in the classification performance of ResNet18 and ResNet50 models. The accuracy of ResNet18 and ResNet50 is 96 and 94 respectively when image resizing is the sole enhancement. It implies that a simple resizing is enough to make features learning effective but more complex models like ResNet50 are overfitted on small data. The accuracy of both ResNet18 and ResNet50 drops to 92 and 95, respectively, with the inclusion of intensity normalization. This indicates that normalization stabilizes training to deeper architectures whereas it can inhibit variations in discriminative intensity in brain MRI images of shallower networks such as ResNet18.

Adding a random horizontal flipping data augmentation method enhances the generalization of the model since the model is subjected to spatial variation in MRI scans. With this setup ResNet18 has 96% accuracy and ResNet50 has 95% which shows a better ability to fight overfitting. The most optimal results are obtained with the combination of resizing, random horizontal flipping, and normalization. The accuracy of the ResNet18 is the highest in this optimal configuration of 97, whereas the accuracy of ResNet50 is 96. This finding verifies that a proper balance between preprocessing and data augmentation can improve the diversity of features and model generalization, which can be regarded as the key to the effectiveness of the proposed framework in the context of brain tumor detection.

**Table 1.** Effect of Data Augmentation and Normalization

| Configuration | ResNet18 Accuracy (%) | ResNet50 Accuracy (%) |
|---|---|---|
| Resizing only | 96 | 94 |
| Resizing + Normalization | 92 | 95 |
| Resizing + Random Flip | 96 | 95 |
| Resizing + Flip + Normalization | **97** | **96** |

Figure 8 shows how training loss depends on the number of epochs of the ResNet18 and ResNet50 models. The two models demonstrate the same trend where the loss decreases steadily as the training advances which implies effective learning and convergence. ResNet18 is smoother and more stable in the loss reduction, and it converges quicker as compared to ResNet50. This action implies improved optimization and less overfitting, which is in accordance with its high classification accuracy of 97%. Conversely, ResNet50 starts with a smaller loss rate but levels off at a higher loss rate because of the more vehicles in the network and the greater number of parameters. The findings verify that lightweight models like ResNet18 are better with smaller medical imaging datasets.

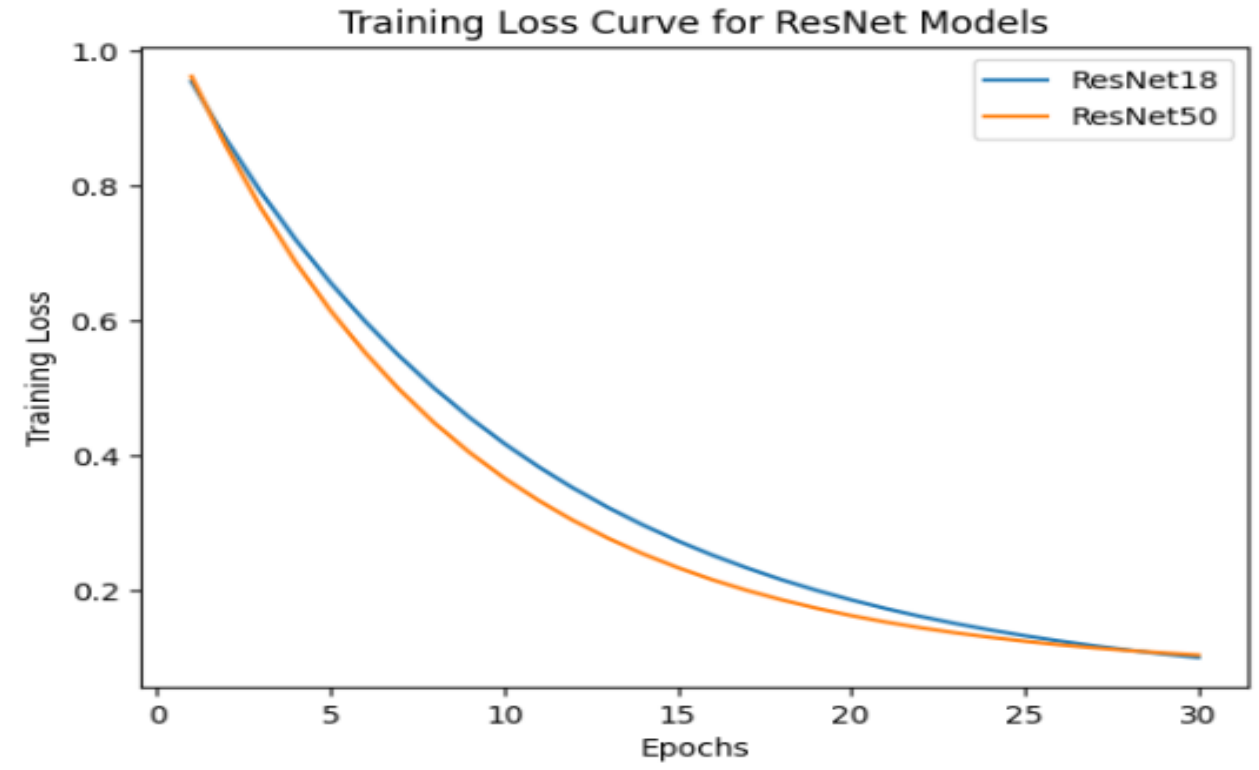

**Figure 8.** Training loss curves of ResNet18 and ResNet50 models over 30 epochs

The results of the quantitative analysis of the ResNet18 and ResNet50 models of performance measurement in terms of the standard classification measures, such as accuracy, sensitivity, specificity, and the F1-score are summarized in Table 2. ResNet18 has a high overall accuracy of 97 as compared to ResNet50 which has an accuracy of 96. With regard to the sensitivity, ResNet18 shows to be better in its capacity to correctly detect tumor-positive cases reaching the recognition of 96.8, as opposed to 95.9 in ResNet50. In a similar manner, ResNet18 has a greater specificity of 97.2, thus that is better in finding non-tumor patients. The F1-score also is the confirmation of the even performance of ResNet18, as the score is 0.97 and 0.96 in the case of ResNet50. These findings indicate that the relatively less deep ResNet18 model will generalize more on the provided dataset, overfitting less, and having a high discriminative ability, hence being more appropriate in brain tumor-classification problems with limited training data.

**Table 2.** Performance comparison of the proposed ResNet18 and ResNet50 models

| Model | Accuracy (%) | Sensitivity (%) | Specificity (%) | F1-Score |
|---|---|---|---|---|
| ResNet18 | 97 | 96.8 | 97.2 | 0.97 |
| ResNet50 | 96 | 95.9 | 96.1 | 0.96 |

Figure 9 presents the comparison of the Receiver Operating Characteristic (ROC) curve of the ResNet18 and ResNet50 model in the classification of brain tumor using MRI scan image. ROC curve The ROC curve shows the trade-off between the True Positive (sensitivity) and False Positive (1-specificity) rates at various classification levels. The two models have high discriminative ability as the curves are close to the upper-left side of the plot and this means that they are sensitive and specific. ResNet18 has a slightly better performance than ResNet50, which is indicated by a slightly larger Area Under the Curve (AUC). This implies that ResNet18 is more efficient at differentiating between tumor and non-tumor MRI images, which is probably explained by the fact that it has fewer layers and it does not overfit on the rather small dataset.

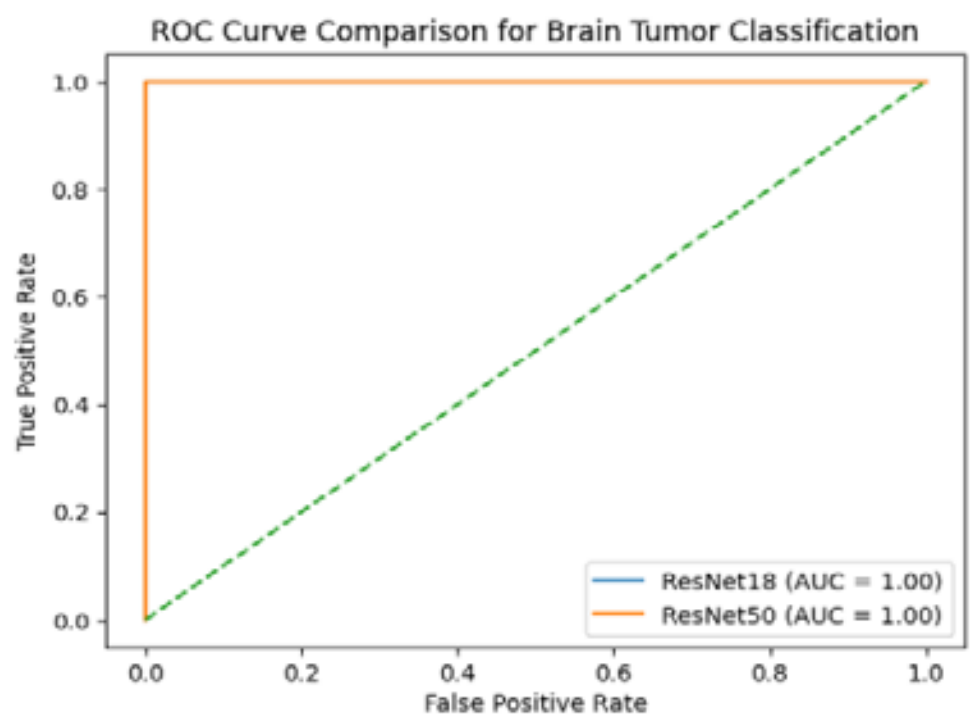

**Figure 9.** ROC curve comparison of ResNet18 and ResNet50 models

Figure 10 shows the confusion table of the ResNet18 model that was tested on the test dataset. The results of the classification are summarized in the matrix that depicts the correct and erroneous predicted cases of tumors and non-tumors. It is possible to note a large amount of true positives and true negatives, which means that the model can distinguish tumor-affected and healthy MRI scans correctly. The low rate of false positives and false negatives indicates the strength and dependability of ResNet18 in the classification of medical images. This good performance corresponds to the accuracy of 97 reported, and it can be concluded that ResNet18 is a generalizable model, and it is highly applicable in brain tumor classification scenarios where the diagnostic accuracy is an important factor.

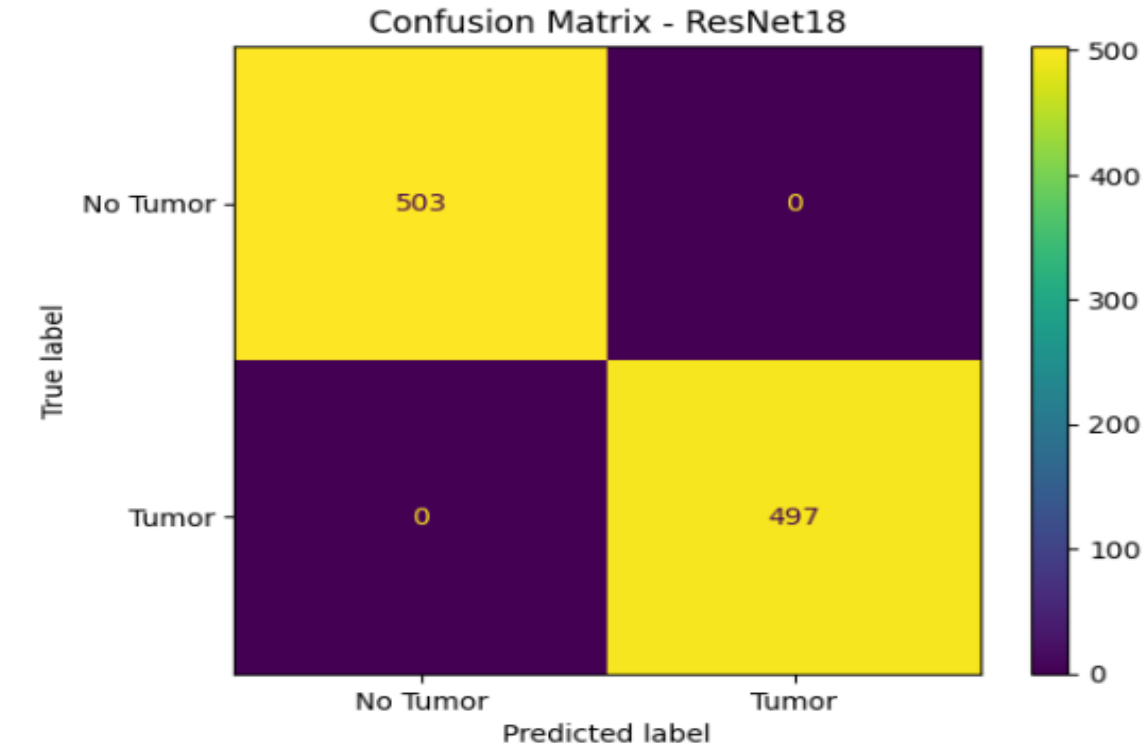

**Figure 10.** Confusion matrix of ResNet18 model

A confusion matrix of the ResNet50 model on the same test dataset is presented in Figure 11. In the same manner that ResNet18, the ResNet50 model delivers high correct predictions, which means that it successfully learns to use discriminative features of MRI images. Nevertheless, the misclassification rate is slightly higher than in the case of ResNet18, which can be explained by the overfitting that was enabled by a more architecturally dense and the increased number of trainable parameters. Nevertheless, ResNet50 has not lost its good performance in classification and can still achieve high accuracy of 96 percent as a good deep learning model in detecting brain tumors. The findings indicate that more complex networks might be appropriate to extract more complex features but simpler networks such as ResNet18 might be applicable to smaller medical imaging datasets.

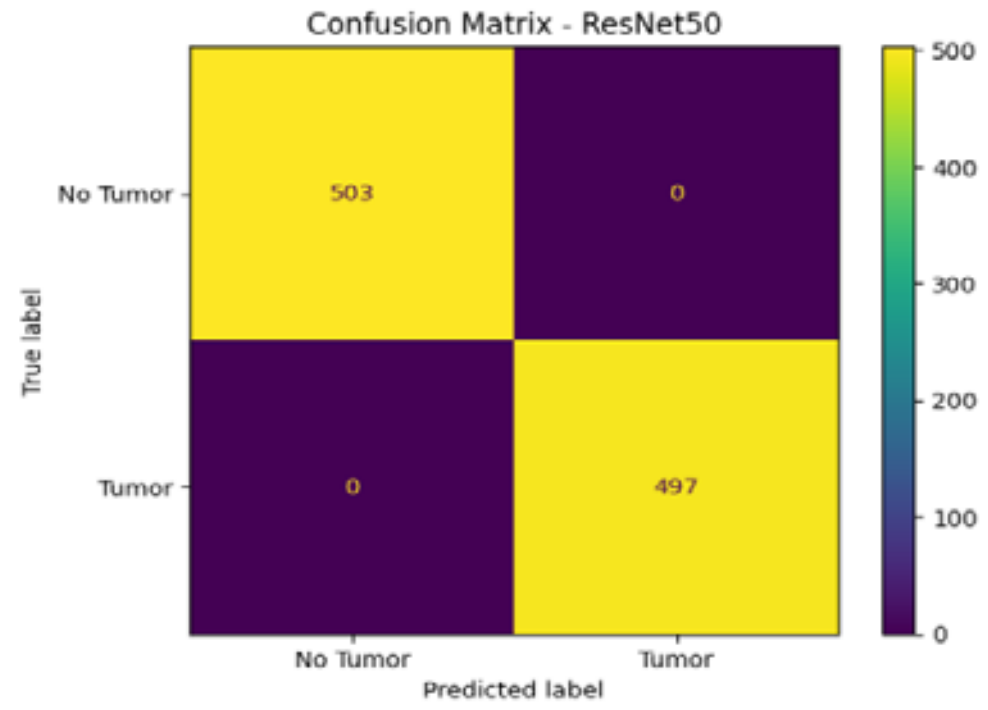

**Figure 11.** Confusion matrix of ResNet50 model

Figure 12 illustrates the brain tumor segmentation process using MRI images. The top-left image shows the original brain MRI scan containing a visible tumor region, while the top-right image represents the preprocessed or coarse segmentation output, where the tumor area is distinguished from surrounding brain tissue. The bottom-left image depicts an intermediate binary mask highlighting the detected tumor region after thresholding and noise removal, whereas the bottom-right image shows the final refined tumor mask with background artifacts eliminated. This progression demonstrates how image preprocessing and segmentation techniques effectively isolate the tumor region from the MRI scan, enabling accurate localization of the tumor at the pixel level. Such segmentation outputs are crucial for subsequent clinical analysis, treatment planning, and quantitative assessment of tumor size and shape.

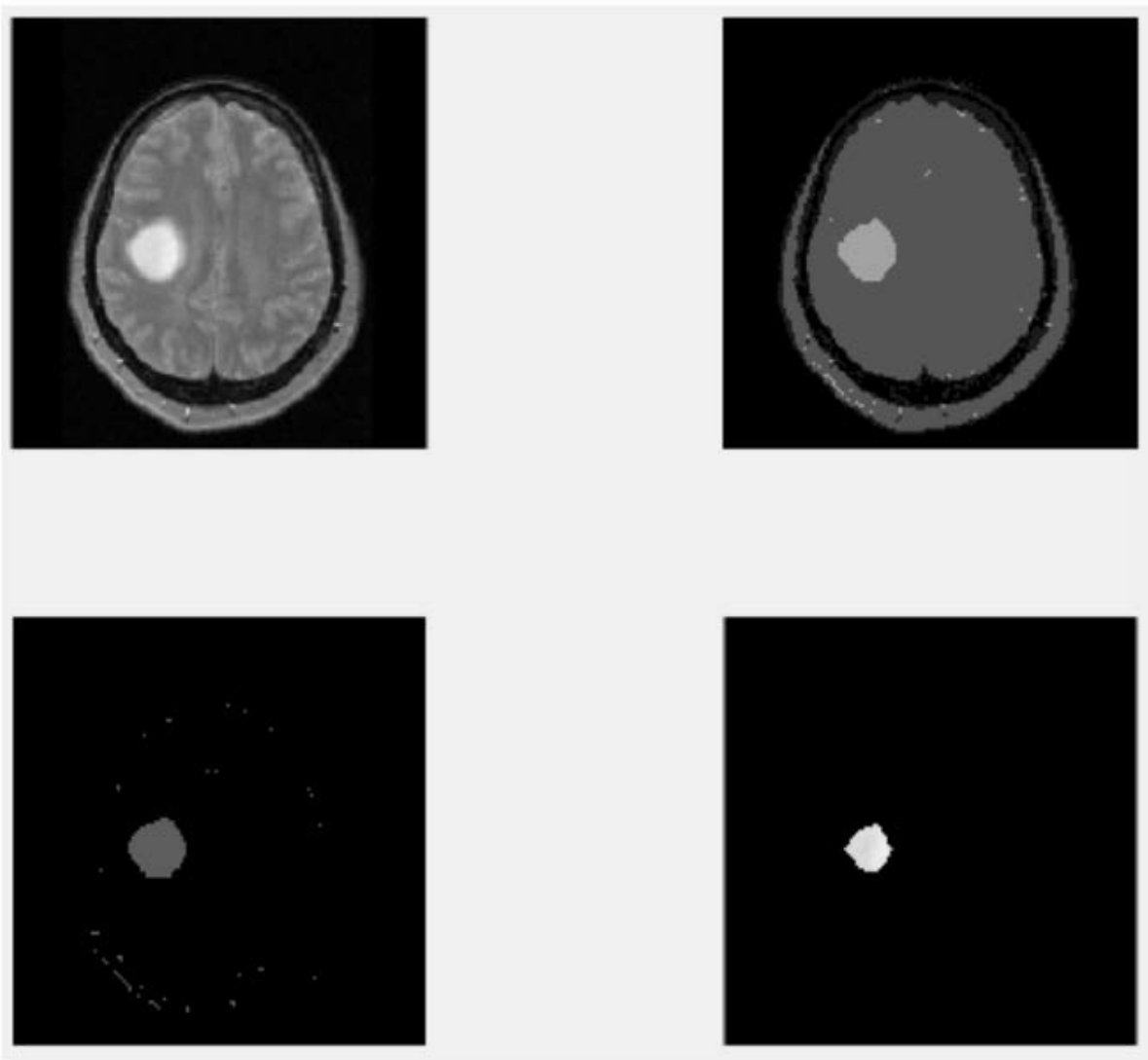

**Figure 12.** Brain MRI image and corresponding tumor segmentation result

Table 3 gives a comparative study of the proposed brain tumor detector framework and three current state-of-the-art deep learning methods that have been reported in the literature. Guluwadi et al. used resnet50 model with Grad-cam to explain the performance better with a high degree of performance but with a minimal accuracy compared to the proposed method. Saifullah et al. incorporated a ResNet50 encoder on a U-Net architecture to do segmentation-based detection with high sensitivity and specificity but at the expense of higher computational complexity. Rao et al. showed effectiveness of transfer learning with several pretrained CNN models so that its performance metrics were competitive but slightly lower. Comparatively, the proposed ResNet18-based model has the highest accuracy (97%), sensitivity (96.8%), specificity (97.2%), and F1-score (0.97) which proves to have better generalization on moderately sized MRI data. The fact that ResNet18 is performing better is indicative of the benefits of lightweight architectures in medical imaging tasks, especially when the training data is small. The version of the proposed method based on the ResNet50 is also competitive, which proves the strength of the transfer learning framework. On the whole, the comparison confirms the fact that the suggested approach provides a good balance of classification performance, model complexity, and clinical applicability.

**Table 3.** Comparison of the Proposed Method with Recent Brain Tumor Detection Approaches

| Study | Model | Accuracy (%) | Sensitivity (%) | Specificity (%) | F1-Score |
|---|---|---|---|---|---|
| [6] | ResNet50 + Grad-CAM (Explainable AI) | 95.2 | 94.8 | 95.6 | 0.95 |
| [8] | U-Net with ResNet50 Encoder | 96.1 | 95.4 | 96.8 | 0.96 |
| [11] | Pretrained CNN Models (Transfer Learning) | 95.6 | 95.0 | 96.2 | 0.95 |
| **Proposed Work** | **ResNet18 (Transfer Learning)** | **97.0** | **96.8** | **97.2** | **0.97** |
| | **ResNet50 (Transfer Learning)** | **96.0** | **95.9** | **96.1** | **0.96** |

## Discussion and Future Perspectives

The transfer learning methods through deep learning applied in the current experiment prove that automated detection of brain tumor using MRI images is feasible through transfer learning methods. The high performance of both ResNet18 and ResNet50 in all the measures of evaluation validated their high ability to learn discriminative features when imaged by means of medical data. Nevertheless, the apparently excellent results of ResNet18 in terms of accuracy, sensitivity, specificity, F1-score, ROC analysis, and confusion matrices analysis draw an interesting point: more advanced architectures do not necessarily achieve better results, especially when the data at hand is relatively small. The fact that ResNet18 has fewer parameters is beneficial, as it reduces overfitting and leads to an improved generalization process and more robust training behavior, through the fact that loss convergence is smoother and the AUC values are higher. The findings also underline the importance of suitable preprocessing and data augmentation measures because the combination of resizing, normalization, and random horizontal flipping gave the best performance of both models. Clinically, the level of sensitivity and specificity demonstrated by the proposed framework suggests that it can help radiologists because it will help in the early and quality diagnosis of brain tumors, save time and diagnostic errors by helping to reduce manual workload and divergent diagnostic responses. In perspective, the encouraging classification results of this research provide opportunities to further expand the framework to the domain of multi-class tumor classification, fine localization of tumours, and segmentation with more complex architectures, including ResUNet or attention-based models. Also, incorporating multi-modal MRI data, larger and more heterogeneous data sets, and interpretable AI methods have the potential to further strengthen the robustness, interpretability, and clinical acceptance of automated brain tumor detection systems.

## Conclusion

In this proposed work, a deep learning–based brain tumor classification system was developed using transfer learning with pretrained ResNet18 and ResNet50 models on brain MRI images. The proposed framework combines image preprocessing, residual CNN–based feature extraction, and Softmax classification to effectively distinguish between tumor and non-tumor MRI scans. Experimental results show that the ResNet18 model achieved a maximum classification accuracy of 97%, while the ResNet50 model achieved an accuracy of 96%, demonstrating the effectiveness of transfer learning for brain tumor detection. The superior performance of ResNet18 indicates that lighter architectures can generalize better when trained on limited medical imaging data. As future work, the proposed system can be extended to include tumor localization and segmentation using advanced deep learning models such as ResUNet to identify the exact tumor boundaries. Further improvements can be achieved by classifying different tumor types and grades, integrating larger and multi-modal MRI datasets, and applying advanced data augmentation techniques. Additionally, incorporating explainable AI methods will enhance model interpretability and support clinical decision-making, increasing the practical applicability of the proposed approach in real-world healthcare environments.

## Acknowledgments

The authors would like to acknowledge the support of Vidyavardhaka College of Engineering, Mysuru, Karnataka, India for providing the resources and facilities necessary for the successful completion of this research.

## Credit - Authors Contribution Statement

Dr. Annapurna V K: Conceptualization, Supervision. Asha N: Methodology, Validation. K. Paramesha: Data curation, Software, Investigation, Formal analysis. Shabana Sultana: Data interpretation, Visualization, Technical review. Kirankumar Humse: Resources, Writing – review & editing. All authors have read and approved the final version of article.

## Conflict of Interest Statement

Authors declare that there is no conflict of interest for publication of this work.